\documentclass[twoside,twocolumn,pra]{revtex4}
\usepackage[T1]{fontenc}
\usepackage{mathptmx} 
\usepackage[scaled=.75]{beramono}  
\usepackage[utf8]{inputenc}
\usepackage{amssymb}
\usepackage{amsthm} 
\usepackage{mathrsfs}
\usepackage{eurosym}
\usepackage{tikz}
\usetikzlibrary{shapes,arrows,patterns} 
\usepackage[europeanresistors,americaninductors]{circuitikz}\ctikzset{bipoles/length=5ex}

\definecolor{gold}{rgb}{1.,.95,0} 

\definecolor{rot}{rgb}{.9,0,0} 

\definecolor{gruen}{rgb}{0,.6,0} 

\definecolor{blau}{rgb}{0,0,.6} 

\usepackage{hyperref,breakurl}

\newtheorem{satz}{Theorem}

\newtheorem{defi}[satz]{Definition}

\newtheorem{rem}[satz]{Remark}
\newtheorem{beisp}[satz]{Example}

\newtheorem{beispe}[satz]{Examples}
\newtheorem{exercise}[satz]{Exercise}

\newtheorem{algorithm}[satz]{Algorithm}

\newenvironment{beispiel} {\begin{beisp}\begin{em}}{\end{em}\hfill{$\square$}\end{beisp}}

\newcommand{\fett}[1]{\mbox{\textbf{\textit{#1}}}}

\begin{document}
\title{Electricity markets regarding the operational flexibility of power plants}

\author{Cem Kiyak}
\author{Andreas de Vries}
\affiliation{South Westphalia University of Applied  Sciences, 
Haldener Stra{\ss}e 182,
D-58095 Hagen, Germany}

\date{2015-01-31}


\begin{abstract}
Electricity market mechanisms designed to steer sustainable generation of electricity
play an important role for the energy transition intended to mitigate climate change.
One of the major problems is to complement volatile renewable energy 
sources by operationally flexible capacity reserves.
In this paper a proposal is given to determine prices on 
electricity markets taking into account
the operational flexibility of power plants, such that the costs of 
long-term capacity reserves can be paid by 
short-term electricity spot markets. For this purpose,
a measure of operational flexibility is introduced enabling to compute 
an inflexibility fee charging each individual power plant 
on a wholesale electricity spot market.
The total sum of inflexibility fees accumulated on the spot markets 
then can be used to finance
a capacity market keeping 
the necessary reserves to warrant grid reliability.
Here each reserve power plant then gets a reliability payment depending 
on its operational flexibility.
The proposal is applied to a small exemplary grid, illustrating its
main idea and also revealing the caveat that 
too high fees paradoxically could create incentives to employ
highly flexible power plants on the spot market rather than to run them as
backup capacity.
	

	\medskip
	\noindent
	\textbf{\href{https://www.aeaweb.org/econlit/jelCodes.php}{JEL}}
	C02, C70, D47, H23, Q21, Q41

	\medskip
	\noindent
	\textbf{\href{http://www.acm.org/class/1998/}{ACM Classification (1998):}}
	J.4 
	
	\medskip
	\noindent
	\textbf{\href{http://dl.acm.org/ccs.cfm}
	{2012 ACM Computing Classification System:}}
Theory of computation $\to$ 
Computational pricing and auctions

	\medskip
	\noindent
	\textbf{\href{http://cdn.elsevier.com/promis_misc/egyclass.pdf}
	{Elsevier Subject Classification for Energy:}}
	13.101 Electricity Markets
\end{abstract}

\maketitle

\section{Introduction}

To mitigate the global climate change
it is commonly agreed that 
greenhouse gas emissions, and in particular emissions of CO$_2$, 
have to be reduced substantially
\cite{
Foster-et-al-2010,IPCC-2007,IPCC-2013,IPCC-2014,Rockstroem-et-al-2009,Rockstroem-et-al-2009-Nature}.
Since 85\% of current primary energy driving global economies 
are due to the combustion of fossil fuels,
and since consumption of fossil fuels accounts for 56.6\% of all anthropogenic 
greenhouse gas emissions, introducing renewable energy sources 
to support all areas of human life plays an essential role
in fighting global warming \cite{Edenhofer-et-al-2012}.
In particular, the generation of electricity by renewables will be
an important step towards this goal, requiring substantial changes
to current grid structures and power plant systems.

If generation and distribution of electricity is to be organized by market principles,
a preeminent challenge of a future electricity market mechanism design is to set effective price signals
to reward the introduction and the use of renewable energy sources 
for the generation of electricity,
and simultaneously to penalize fossil fuel power plants.
However, the physical requirements of electricity grids
and the necessities of public life in present societies 
impose special restrictions on
electricity markets. 
In particular, a necessary condition for
grid stability is the reliability of electricity generation
and the immediate equality of supply and demand at any instant of time.
It is expected that the biggest contribution of renewable energy sources in electricity grids
will come from wind turbines and photovoltaic cells \cite{Agora-2013},
both producing electricity only with high volatility.
Their widespread installation therefore would challenge the reliability of electricity 
supply and thus the stability of the grids.
Lacking sufficiently large storages for electricity,
to warrant reliability in grids with volatile energy sources
power plants with high operational flexibility are required
as a power reserve standing in 
in cases of sudden scarcity of electricity supply or of blackouts.
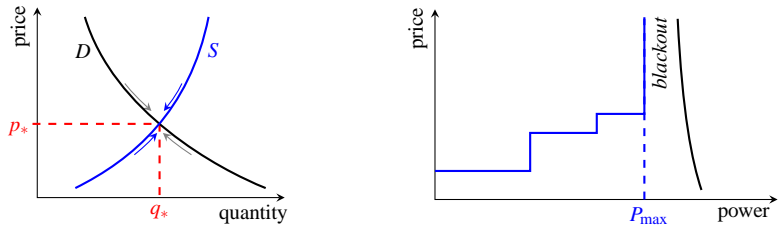
\begin{figure*}[htp]
\centering
\begin{footnotesize}
\begin{tikzpicture}[>=stealth]
	\draw[<->]
		(0ex,20ex) -- node[near start,rotate=90,anchor=south west] {price}
		(0ex, 0ex) -- node[very near end,anchor=north] {quantity}
		(26ex,0ex);
	\draw[dashed,red,thick] (-.5ex,7.75ex) -- (12.85ex,7.75ex) -- (12.85ex,-.5ex);
	\draw[->,gray] (9.2ex,12.0ex) .. controls (10.2ex,10.8ex) .. (12.0ex,9.1ex);	
	\draw[->,blue] (15.2ex,12.0ex) .. controls (14.58ex,10.8ex) .. (13.3ex,9.1ex);	
	\draw[->,gray] (16.2ex,4.5ex) .. controls (14.6ex,5.6ex) .. (13.2ex,6.8ex);	
	\draw[->,blue] (10.2ex,4.5ex) .. controls (11.5ex,5.6ex) .. (12.6ex,6.8ex);	
	\node[anchor=east,red] at (0ex,7.25ex) {$p_*$};
	\node[anchor=north,red] at (12.9ex,0ex) {$q_*$};
	\draw[thick] (5ex,19ex) .. controls (8ex,9ex) and (19ex,3ex) .. node[very near start,anchor=east] {$D$} (24ex,1ex);
	\draw[thick,blue] (18ex,19ex) .. controls (16ex,9ex) and (10ex,4ex) .. node[very near start,anchor=west] {$S$} (4ex,1ex);
\end{tikzpicture}
\hspace*{10ex}
\begin{tikzpicture}[>=stealth]
	\draw[<->]
		(0ex,20ex) -- node[near start,rotate=90,anchor=south west] {price}
		(0ex,0ex) -- 
		(36ex,0ex);
	\node[anchor=north,blue] at (22.5ex,0ex) {$P_{\max}$};
	\node[anchor=north east] at (36ex,0ex) {power};
	\draw[thick, blue] 
		(0ex,3ex) -- (10ex,3ex) -- (10ex,7ex) -- (17ex,7ex) -- (17ex, 9ex) -- (22ex,9ex) -- (22ex,19ex);
	\draw[dashed,blue,thick] (22ex,0ex) -- (22ex,20ex);
	\draw[thick] (25.5ex,19ex) .. controls (26ex,9ex) and (27ex,4ex) .. (28ex,1ex);
	\node[rotate=90, anchor=east] at (23.6ex,20ex) {\em blackout};
\end{tikzpicture}
\end{footnotesize}
\caption{\label{fig-blackout}\footnotesize
	A perfect efficient market where any demand $D$ meets supply $S$
	at a certain equilibrium price $p_*$ and quantity $q_*$ (left hand).
	On an electricity spot market, a 
	blackout is a market failure due to inelastic demand and supply,
	with the supply curve given by the merit order of the power plant
	system (right hand).
	Here a (“rolling”) blackout occurs if the demand is higher 
	than the total maximum power 
	$P_{\max}$ of all power plants \cite{Cramton-et-al-2013}.
	Increasing the demand-side inelasticity, e.g., by smart grids,
	could remove the problem on the long run, but in the short run
	electricity markets require capacity reserves which are not demanded
	for most of the time.
}
\end{figure*}
Cramton and Ockenfels \cite{Cramton-Ockenfels-2012}
proved the “missing money” theorem stating that,
in a competitive electricity market, prices are always too low to
pay for adequate capacity. 
In fact, present electricity markets are not perfect efficient markets
since both supply and demand are price inelastic, see
Figure~\ref{fig-blackout}.
Future increase of demand elasticity, for instance by smart grids,
would relax the difficulties to a certain degree, 
but inelasticity on the supply-side could only be removed by
capacity reserves or huge electricity storages.
The first option, however, requires long-term plannings at a magnitude
of decades, whereas the second option is technologically not
realizable to date. For more details see \cite{Cramton-et-al-2013}.

Besides these theoretical arguments there also exist empirical 
clues to doubt that current electricity 
markets encourage investments in operationally flexible power plants or 
in the provision
of power reserves for cases of emergency or maintenance
\cite{Agora-2013-capacity-market,
Ecke-et-al-2013-en,EWI-2012,Steuwer-2013}.
Several solutions to this problem have been
proposed recently to complement the present “energy-only” markets, 
ranging from 
separate capacity markets which trade backup capacity,
to strategic capacity reserves usually settled by long-term contracts 
with national agencies
\cite{Agora-2013-capacity-market,Cramton-Ockenfels-2012,EWI-2012,Gawel-Purkus-2013,Growatsch-et-al-2013,Matthes-et-al-2012,Schoenberg-2015,Steuwer-2013}.

The main goal of this paper is to propose a solution for the economic 
problem to finance the necessary capacity reserves guaranteeing grid 
stability by market principles.
One necessary property of a power plant for being part of a capacity reserve is a fast guaranteed operational flexibility.
In our opinion the main problem of current market mechanism designs 
is the fact that market prices do not regard operational flexibility, 
being determined solely by the marginal costs of 
electricity generation. Thus the costs of operational inflexibility 
are market externalities 
\cite[§14]{Bofinger-2007}, \cite[p 125]{Krugman-Wells-2006}
and reduce welfare.
By contrast, a sustainable electricity market mechanism design should induce market prices
which take into account both the direct variable production costs and the 
external ecological costs of 
electricity production, but also the costs caused by operational inflexibility of
each individual power plant.
Due to emissions trading \cite[§15.4]{Stroebele-et-al-2012}, 
the first two cost factors are 
already priced in as marginal costs
on present electricity spot markets,
but operational flexibility does not play a role for the 
determination of the spot market prices to date.
To internalize it into price calculation, at first we define 
measure for the operational flexibility of a given power plant. 
This measure then can be used
to compute an inflexibility fee for each power plant.
The total of these inflexibility fees then can serve to 
pay power reserves provided by some given capacity mechanism.

This paper is organized as follows. In section 
\ref{sec-definition-operational-flexibility} 
a general class of functions measuring the operational flexibility 
of a power pant 
in dependence to its guaranteed start-up time is defined.
In section \ref{sec-inflexibility-fee} the effect of the inflexibility fee
on the offer price of a power plant on an electricity spot market 
is calculated
and demonstrated by a prototypical exemplary “toy” grid in Example 
\ref{beisp-toy-grid}.
A way how the accumulated inflexibility fees then can be used to finance 
a capacity mechanism
is described in section \ref{sec-capacity-reserve}, 
before a short discussion concludes the paper.

\section{A measure of the operational flexibility of a power plant}
\label{sec-definition-operational-flexibility}%
We stipulate that the operational flexibility of a power plant 
depends on its
\emph{guaranteed start-up time}\index{guaranteed start-up time} 
$t_s \in [0,\infty)$
which is defined as the time that a power plant requires to supply a guaranteed
power of electricity.
Moreover, we claim that the measure should be a pure number expressing a degree
of flexibility ranging from $0$ to $1$, 
with the property that the longer the guaranteed start-up time 
the smaller the value of flexibility.
Consequently, we define a general 
\emph{measure of operational flexibility}%
\index{measure of operational flexibility}\index{operational flexibility}\index{flexibility}
to be a strictly monotonically decreasing function $\varphi: [0,\infty) \to [0,1]$
of a single variable satisfying the limit behavior
\begin{equation}
	\varphi(x) \to 1 \quad \mbox{as \ $x \to 0$,}
	\quad \mbox{and} \quad
	\varphi(x) \to 0 \quad \mbox{as \ $x \to \infty$}
	.
	\label{eq-operational-flexibility-norm-condition}
\end{equation}
Here the variable $x$ represents the starting time 
of the power plant, measured in hours [h].
A simple example of such a measure is the differentiable function
\begin{equation}
	\varphi(x)
	= \frac{1}{x+1}
	.
	\quad 
	\begin{tabular}{c}
	\begin{tikzpicture}[>=stealth, domain=0:3, yscale=1.2]
		\draw[->,darkgray] (-0.2,0) -- (3.2,0) node[below] {$x$};
		\draw[->,darkgray] (0,-0.2) -- (0,1.2) node[left] {$\varphi(x)$};
		\draw 
		plot (\x,{1/(\x + 1)}); 
	\end{tikzpicture}
	\end{tabular}
	\label{eq-operational-flexibility}
\end{equation}
In the sequel we will use this function to measure the operational flexibility of
a given power plant.
In Table \ref{tab-operational-flexibility-measures} there are listed
guaranteed start-up times $t_s$ and the respective flexibility measures
$\varphi(t_s)$ for some typical power plants.
\begin{table}[htp]
\centering
\begin{footnotesize}
\begin{tabular}{|l|c|c||c|}
	\hline
	 & \textbf{Guaranteed} & & 	\textbf{Marginal} \\	
	\textbf{Power Plant} & \textbf{Start-Up} & $\mathbf{\varphi}$ &
	\textbf{Costs $\fett{p}^{\mathbf{mc}}$}
	\\
	 & \textbf{Time [h]} & & \textbf{[\euro/MWh]}
	\\ \hline
	wind turbine                & $\infty$ & .000 & --- \\ 
	hydroelectric power station &  0.02    & .979 & --- \\ 
	gas turbine                 &  0.12    & .893 & 90  \\ 
	cogeneration plant (CHP)    &  0.17    & .855 & --- \\ 
	combined cycle gas turbine  &  5       & .167 & 50  \\ 
	hard coal power plant       &  6       & .143 & 60 \\ 
	lignite power plant         &  9       & .100 & 40 \\ 
	nuclear power plant         & 50       & .020 & \,\, 5 \\ 
	\hline
\end{tabular}
\end{footnotesize}
\caption{\label{tab-operational-flexibility-measures}\footnotesize
	Exemplary cold start-up times and their respective operational flexibility measures,
	as well as exemplary marginal costs (without emissions trading).
	The values are typical for current German electricity markets.
	Data from \cite[p 71]{Grimm-2007-Diss} (start-up times)
	and 
	\cite[pp 13, 19]{Heitmann-2010}, 
	\cite[p 3]{von-Roon-Huck-2010} (marginal costs).
}
\end{table}
Note that a wind turbine is assigned a vanishing operational flexibility,
since due to the volatility of winds 
a predetermined amount of energy by a wind turbine cannot be guaranteed
at a given future instant.
The highest operational flexibilities are exposed by hydroelectric power stations
and modern gas turbines.

\section{Fees on operational inflexibility}
\label{sec-inflexibility-fee}%
On an wholesale electricity market,
each participating power plant operator offers electric power with a sell bid 
for each of its power plant.
The market maker collects all these sell bids and determines the market-clearing price
in accordance to the buy bids and the merit order
\cite{Das-et-al-2001,EPEX-2014,Wilson-1985,Satterthwaite-Williams-1989},
for a theoretical introduction see also \cite[§6.5, §7.4.5]{Fudenberg-Tirole-1991}.	
Our main idea now is to rise a fee for operational inflexibility on each
power plant, its amount being calculated by the operational flexibility $\varphi$
as part of a factor to a given market-wide reference level.
In consequence, the offer price of each power plant must
take it into account its operational flexibility.

To be more precise, let
$p_i^{\mathrm{mc}}$ denote the marginal offer price per energy quantity of the power plant
regarding only the marginal costs,
including the variable costs of production and emissions trade certificates;
this is the price which would be offered for the power plant on a current wholesale spot market
\cite{von-Roon-Huck-2010}.
Assume moreover that all power plants participating at the spot market are uniquely numbered 
by the indices $i$ $=$ $1$, $2$, \ldots, $n$.
The spot market offer price $p_i$ of plant $i$ taking into account its operational flexibility
$\varphi_i$ then is calculated by the formula
\begin{equation}
	\fbox{$
	p_i
	= p_i^{\mathrm{mc}} + (1 - \varphi_i) \ p_0
	.
	$} 
	\label{eq-price-regarding-operational-flexibility}
\end{equation}
Here $p_0$ denotes a market-wide constant reference level price, set by the market
authority.
It therefore is a political or regulatory quantity,
not a market-inherent value or immediately economically deducible.
It is arbitrary in principle, but the higher
its amount the heavier the effect of operational flexibility on the final
spot market price.
\begin{table*}[htp]
\centering
\begin{footnotesize}
\begin{tabular}{|l|c|r||r|r|r|}
	\hline
	 & & & \textbf{Marginal Price} & \multicolumn{2}{c|}{\textbf{Offer Price}} \\
	\textbf{Power Plant} 
	 & $\mathbf{\varphi}$ 
	 & $\mathbf{1 - \varphi}$ 
	 & \multicolumn{1}{c|}{$\fett{p}^{\mathbf{mc}}_i$}
	 & \multicolumn{2}{c|}{$\fett{p}_i$ \textbf{[\euro/MWh]}}
	\\ \cline{5-6}
	 & & 
	 & \multicolumn{1}{c|}{\textbf{[\euro/MWh]}}
	 & \multicolumn{1}{c|}{$\fett{p}_0$ = \textbf{10}}
	 & \multicolumn{1}{c|}{$\fett{p}_0$ = \textbf{70}}
	\\ \hline
	wind turbine                
		& .000 & $1.000$ &  $1$ \qquad \ & $11$ \hspace*{1.2ex} &  $71$ \hspace*{1.2ex}
	\\
	hydroelectric power station 
		& .980 & $.020$ &  $1$ \qquad \ &   $1$ \hspace*{1.2ex} &   $2$ \hspace*{1.2ex}
	\\
	gas turbine                 
		& .893 & $.107$ & $90$ \qquad \ &  $91$ \hspace*{1.2ex} &  $98$ \hspace*{1.2ex}
	\\
	cogeneration plant (CHP)    
		& .855 & $.145$ & $50$ \qquad \ &  $51$ \hspace*{1.2ex} &  $60$ \hspace*{1.2ex}
	\\
	combined cycle gas turbine  
		& .167 & $.833$ & $50$ \qquad \ &  $58$ \hspace*{1.2ex} & $108$ \hspace*{1.2ex}
	\\
	hard coal power plant       
		& .143 & $.857$ & $60$ \qquad \ &  $69$ \hspace*{1.2ex} & $120$ \hspace*{1.2ex}
	\\
	lignite power plant         
		& .100 & $.900$ & $40$ \qquad \ &  $49$ \hspace*{1.2ex} & $103$ \hspace*{1.2ex}
	\\
	nuclear power plant         
		& .020 & $.980$ &  $5$ \qquad \ &  $15$ \hspace*{1.2ex} &  $74$ \hspace*{1.2ex}
	\\
	\hline
\end{tabular}
\end{footnotesize}
\caption{\label{tab-prices-regarding-operational-flexibility}\footnotesize
	Operational flexibilities, as given by the exemplary data 
	of Table \ref{tab-operational-flexibility-measures},
	and the resulting offer price differences with respect to the reference level prices
	$p_0 = 10$ \euro$/$MWh and $p_0 = 70$ \euro$/$MWh.
}
\end{table*}
It should be high enough to signal effective incentives to introduce and use
operationally flexible power plants for scarcity situations and black-outs,
but it must be low enough to avoid a too radical change of the merit order
such that too many flexible power plants are operational on the spot market
and thus unavailable for a capacity reserve
(see Figure \ref{fig-operational-flexibility-example}).

\begin{beispiel}
	\label{beisp-toy-grid}%
	Consider a small examplary grid
	(called “toy grid” in the sequel)
	consisting of the eight
	power plants listed in
	Table \ref{tab-operational-flexibility-measures}.
	The prices resulting from the respective inflexibility fees 
	in dependence to different reference level prices
	$p_0$ are listed in 
	Table \ref{tab-prices-regarding-operational-flexibility}.
	If the reference level price is low (here $p_0$ $=$ $10$ \euro$/$MWh),
	the modified offer prices do not change the merit order of the power plant system,
	whereas a sufficiently high reference level price 
	(e.g., $p_0 = 70$ \euro$/$MWh) 
	changes it,
	as is depicted in Figure \ref{fig-operational-flexibility-example}.
\begin{figure*}[htp]
\centering
\begin{footnotesize}
\begin{tikzpicture}[>=stealth,baseline]
	\draw[fill=cyan]      ( 0ex,40ex) rectangle node[rotate=90,anchor=west] {hydro}
        ( 5ex,40.25ex);
	\draw[fill=green]     ( 5ex,40ex) rectangle node[rotate=90,anchor=west] {wind}
        (10ex,40.25ex);
	\draw[fill=gold]      (10ex,40ex) rectangle node[rotate=90,anchor=west] {nuclear}
        (15ex,41.25ex);
	\draw[fill=brown]     (15ex,40ex) rectangle node[rotate=90] {lignite}     (20ex,50.00ex);
	\draw[fill=lightgray] (20ex,40ex) rectangle node[rotate=90] {CHP}         (25ex,52.50ex);
	\draw[gray,fill=cyan]      (25ex,40ex) rectangle 
        (30ex,52.50ex);
	\draw node[rotate=90] at (27.5ex,50ex) {combined cycle gas};
	\draw[fill=lightgray] (30ex,40ex) rectangle node[rotate=90] {hard coal}   (35ex,55.00ex);
	\draw[fill=teal!50]   (35ex,40ex) rectangle node[rotate=90] {gas turbine} (40ex,62.50ex);
	\draw[dashed,red,thick] (-.5ex,52.50ex) node[anchor=south west] {$p_*$}
		-- (23.80ex,52.50ex) -- (23.80ex,39.5ex) node[anchor=north] {$q_*$};
	\draw[<->]
		(0ex,64ex) -- node[very near start,rotate=0,anchor=south west] {\euro$/$MWh}
		(0ex,40ex) -- node[very near end,anchor=north west] {MW}
		(45ex,40ex);
	\draw[<->]
		(60ex,64ex) -- node[very near start,rotate=0,anchor=south west] {\euro$/$MWh}
		(60ex,40ex) -- node[very near end,anchor=north west] {MW}
		(105ex,40ex);
	\draw[dashed,opacity=.5,fill=cyan]      (60ex,40ex) rectangle  (65ex,40.25ex);
	\draw[dashed,opacity=.5,fill=green]     (65ex,40ex) rectangle  (70ex,40.25ex);
	\draw[dashed,opacity=.5,fill=gold]      (70ex,40ex) rectangle  (75ex,41.25ex);
	\draw[dashed,opacity=.5,fill=brown]     (75ex,40ex) rectangle  (80ex,50.00ex);
	\draw[dashed,opacity=.5,fill=lightgray] (80ex,40ex) rectangle  (85ex,52.50ex);
	\draw[dashed,opacity=.5,fill=cyan]      (85ex,40ex) rectangle  (90ex,52.50ex);
	\draw[dashed,opacity=.5,fill=lightgray] (90ex,40ex) rectangle  (95ex,55.00ex);
	\draw[dashed,opacity=.5,fill=teal]      (95ex,40ex) rectangle  (100ex,62.50ex);
	\draw (60ex,40ex) rectangle node[rotate=90,anchor=west] {hydro}    (65ex,40.30ex);
	\draw (65ex,40ex) rectangle node[rotate=90,anchor=west] {\ wind}   (70ex,42.75ex);
	\draw (70ex,40ex) rectangle node[rotate=90,anchor=west] {\ \ nuclear} (75ex,43.70ex);
	\draw (75ex,40ex) rectangle node[rotate=90] {lignite}              (80ex,52.25ex);
	\draw (80ex,40ex) rectangle node[rotate=90] {CHP}                  (85ex,52.86ex);
	\draw (85ex,40ex) rectangle 
		(90ex,54.58ex);
	\draw node[rotate=90] at (87.5ex,50ex) {combined cycle gas};
	\draw (90ex,40ex) rectangle node[rotate=90] {hard coal}            (95ex,57.14ex);
	\draw (95ex,40ex) rectangle node[rotate=90] {gas turbine}          (100ex,62.77ex);
	\draw[dashed,red,thick] (59.5ex,52.86ex) node[anchor=south west] {$p_*$}
		-- (83.80ex,52.86ex) -- (83.80ex,39.5ex) node[anchor=north] {$q_*$};
	\draw[->] (45.0ex,52.0ex)
		--  node[anchor=south] {$p_0 = 10$ \euro/MWh} (55.0ex,52.0ex);
	\draw[fill=cyan]      ( 0ex, 0ex) rectangle node[rotate=90,anchor=west] {hydro}
        ( 5ex, 0.25ex);
	\draw[fill=green]     ( 5ex, 0ex) rectangle node[rotate=90,anchor=west] {wind}
        (10ex, 0.25ex);
	\draw[fill=gold]      (10ex, 0ex) rectangle node[rotate=90,anchor=west] {nuclear}
        (15ex, 1.25ex);
	\draw[fill=brown]     (15ex, 0ex) rectangle node[rotate=90] {lignite}     (20ex,10.00ex);
	\draw[fill=lightgray] (20ex, 0ex) rectangle node[rotate=90] {CHP}         (25ex,12.50ex);
	\draw[gray,fill=cyan]      (25ex, 0ex) rectangle 
        (30ex,12.50ex);
	\draw node[rotate=90] at (27.5ex,10ex) {combined cycle gas};
	\draw[fill=lightgray] (30ex, 0ex) rectangle node[rotate=90] {hard coal}   (35ex,15.00ex);
	\draw[fill=teal!50]   (35ex, 0ex) rectangle node[rotate=90] {gas turbine} (40ex,22.50ex);
	\draw[dashed,red,thick] (-.5ex,12.50ex) node[anchor=south west] {$p_*$}
		-- (23.80ex,12.50ex) -- (23.80ex,-.5ex) node[anchor=north] {$q_*$};
	\draw[<->]
		(0ex,33ex) -- node[very near start,rotate=0,anchor=south west] {\euro$/$MWh}
		(0ex, 0ex) -- node[very near end,anchor=north west] {MW}
		(45ex, 0ex);
	\draw[<->]
		(60ex,33ex) -- node[very near start,rotate=0,anchor=south west] {\euro$/$MWh}
		(60ex, 0ex) -- node[very near end,anchor=north west] {MW}
		(105ex,0ex);
	\draw[dashed,opacity=.5,fill=cyan]      (60ex,0ex) rectangle  (65ex, 0.25ex);
	\draw[dashed,opacity=.5,fill=lightgray] (65ex,0ex) rectangle  (70ex,12.50ex);
	\draw[dashed,opacity=.5,fill=green]     (70ex,0ex) rectangle  (75ex, 0.25ex);
	\draw[dashed,opacity=.5,fill=gold]      (75ex,0ex) rectangle  (80ex, 1.25ex);
	\draw[dashed,opacity=.5,fill=teal]      (80ex,0ex) rectangle  (85ex,23.78ex);
	\draw[dashed,opacity=.5,fill=brown]     (85ex,0ex) rectangle  (90ex,10.00ex);
	\draw[dashed,opacity=.5,fill=cyan]      (90ex,0ex) rectangle  (95ex,12.50ex);
	\draw[dashed,opacity=.5,fill=lightgray] (95ex,0ex) rectangle  (100ex,15.00ex);
	\draw (60ex,0ex) rectangle node[rotate=90,anchor=west] {hydro}  (65ex, 0.59ex);
	\draw (65ex,0ex) rectangle node[rotate=90] {CHP}                (70ex,15.04ex);
	\draw (70ex,0ex) rectangle node[rotate=90] {wind}               (75ex,17.75ex);
	\draw (75ex,0ex) rectangle node[rotate=90] {nuclear}            (80ex,18.06ex);
	\draw (80ex,0ex) rectangle node[rotate=90] {gas turbine}        (85ex,24.38ex);
	\draw (85ex,0ex) rectangle node[rotate=90] {\ lignite}          (90ex,25.75ex);
	\draw (90ex,0ex) rectangle node[rotate=90] {combined cycle gas} (95ex,27.08ex);
	\draw (95ex,0ex) rectangle node[rotate=90] {hard coal}          (100ex,30.00ex);
	\draw[dashed,red,thick] (59.5ex,24.38ex) node[anchor=south west] {$p_*$}
		-- (83.80ex,24.38ex) -- (83.80ex,-.5ex) node[anchor=north] {$q_*$};
	\draw[->] (45.0ex,16.5ex)
		--  node[anchor=south] {$p_0 = 70$ \euro/MWh} (55.0ex,16.5ex);
\end{tikzpicture}
\end{footnotesize}
\caption{\label{fig-operational-flexibility-example}\footnotesize
	Effect of the operational inflexibility fee on the 
	price $p_*$ clearing the market consisting of the power plant system 
	in Table \ref{tab-prices-regarding-operational-flexibility},
	neglecting operational flexibility (left) and regarding it (right).
	The reference level price are assumed as $p_0 = 10$ \euro/MWh 
	and $p_0 = 70$ \euro/MWh, respectively.
	For a given demand $q_*$ of electric power, the market-clearing spot price
	increases more or less slightly, depending on $p_0$.
	For a high operational inflexibility fee, as in the second case, 
	the merit order is changed.
}
\end{figure*}
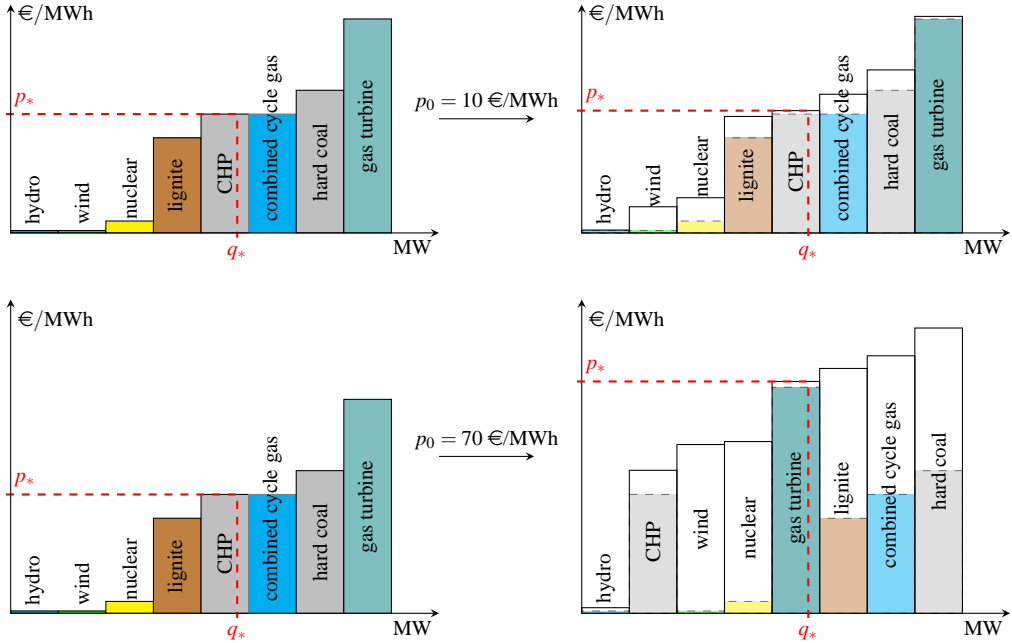
	In our toy grid we can recognize  that, 
	if the amount of $p_0$ is too high,
	the effect may be even counterproductive since the flexible gas turbine
	is in the money and thus operating at a normal quantity demand, 
	leaving no power plant as a capacity reserve. 
	In case of a sudden scarcity or of a blackout, 
	the grid then would perform worse
	than with the original merit order.
	
	Moreover we observe that the higher the reference level price $p_0$, 
	the higher the spot market price.
	The amounts, however, are not related to each other in a linear manner, 
	but depend discontinously on the changes of the merit order.
	The total amount of inflexibility fees, at last, is directly calculated to
	be either 48.4 \euro$/$MWh in case of $p_0$ $=$ 10 \euro$/$MWh,
	or 339 \euro$/$MWh in case of $p_0$ $=$ 70 \euro$/$MWh.
	
	We finally note that for the demanded quantity $q_*$ depicted in the scenarios
	in Figure \ref{fig-operational-flexibility-example},
	only five power plants are operational. Depending on the reference level price
	the realized profit then is given by the following tables.
	
	\noindent
	\begin{footnotesize}
	\[ 
		\begin{tabular}{|l|c|}
			\hline
			\multicolumn{2}{|c|}{\textbf{$\mathbf{\fett{p}_0 = 10}$ \euro/MWh}}
			\\ \hline
			\textbf{Power plant} & \textbf{Profit for} $\mathbf{\fett{q}_*}$
			\\ \hline
			wind turbine  & 40 \euro$/$MWh \\
			hydroelectric & 50 \euro$/$MWh \\
			CHP           & \ \,0 \euro$/$MWh \\
			lignite       & \ \,2 \euro$/$MWh \\
			nuclear       & 37 \euro$/$MWh \\
			\hline
		\end{tabular}
		\quad
		\begin{tabular}{|l|c|}
			\hline
			\multicolumn{2}{|c|}{\textbf{$\mathbf{\fett{p}_0 = 70}$ \euro/MWh}}
			\\ \hline
			\textbf{Power plant} & \textbf{Profit for} $\mathbf{\fett{q}_*}$
			\\ \hline
			wind turbine  & 27 \euro$/$MWh \\
			hydroelectric & 95 \euro$/$MWh \\
			CHP           & 37 \euro$/$MWh \\
			gas turbine   & \ \,0 \euro$/$MWh \\
			nuclear       & 24 \euro$/$MWh \\
			\hline
		\end{tabular}
		\label{eq-toy-grid-profits}
	\] 
	\end{footnotesize}
	Assume for simplicity that the demand remains constantly at $q_*$ during a certain
	hour and that all power plants yield the same power of $5$ MW, say,
	and let be $q_* = 25$ MW be the demanded electrical power for the hour considered
	(such that the consumed electricity energy during this period is $E = 25$ MWh).
	Then with Table \ref{tab-prices-regarding-operational-flexibility}
	the total of the inflexibility fees for the five operational power plants
	in the money 
	has the amount of
	\begin{equation}
		C_f
		= (10 + 1 + 1 + 9 + 10) \cdot 5 
		= 205 \mbox{ \euro$/$h,}
	\end{equation}
	at a reference level price $p_0 = 10$ \euro$/$MWh,
	and
	\begin{equation}
		C_f
		= (70 + 1 + 8 + 10 + 69) \cdot 5 
		= 790 \mbox{ \euro$/$h,}
	\end{equation}
	at a reference level price $p_0 = 70$ \euro$/$MWh.
	The total fee then can be distributed to the power plants participating 
	at a capacity mechanism, paying their time of reliability. 
\end{beispiel}

The toy grid in Example \ref{beisp-toy-grid} demonstrates the possible
direct consequences of the inflexibility fee to the wholesale electricity market.
In essence, by Equation (\ref{eq-price-regarding-operational-flexibility})
a power plant with a low operational flexibility
is penalized more than one with a high operational flexibility.
In the limit case that all power plants participating on the spot market
are equally operationally flexible, i.e., $\varphi_i$ $=$ const,
all sell bids are raised by the same amount and the merit order cannot change.
On the other hand, if the power plants have different operational flexibilities
and the reference price level $p_0$ is chosen too high,
the merit order changes the merit order such that all flexible power plants
are operational on the spot market, such that no power plant is left for
the capacity reserve necessary to warrant grid reliability.

The total amount of inflexibility fees paid for each power plants 
participating the spot market now is available for a capacity mechanism,
as described in the following section.

\section{Accumulated inflexibility fees paying capacity reserves}
\label{sec-capacity-reserve}%
A power plant serving as a power reserve for periods of scarcity or blackouts
should have fast and guaranteed start-up times, i.e.,
should be operationally flexible to a high degree.
There exist several proposed capacity mechanisms,
for instance capacity markets
or a strategic reserve determined by a grid agency.
In either of these approaches,
we therefore require a power plant to offer capacity reserves
to have a high operational flexibility $\varphi$, say
\begin{equation}
	\varphi > \frac12
	.
	\label{eq-capacity-market-flexibility-requirement}
\end{equation}
This value means that the guaranteed start-up time of 
a power plant participating the capacity mechanism
must be less than one hour.
A further natural requirement is that a power plant offering its reliability 
on the capacity market cannot participate on the spot market.

Assume then that there are $k$ power plants participating on the capacity market,
each one established with a unique index $i$ $=$ $1$, \ldots, $k$.
Let $\varphi_i$ and $P_i$ 
denote the operational flexibility and the capacity (measured in MW) 
of power plant $i$, respectively,
and let $C_f$ be the total of inflexibility fees accumulated 
on the spot market in a certain past period, say, the day before.
It has the dimension currency per time, for instance \euro$/$h.
Then the \emph{reliability payment}\index{reliability payment} $\rho_i$
for power plant $i$ in that period is defined as
\begin{equation}
	\rho_i
	= \frac{\varphi_i \, P_i}{P_{\mathrm{flex}}}
	\ C_f
	\quad \left[\frac{\mbox{\euro}}{\mbox{h}}\right]
	\qquad \mbox{where} \qquad
	P_{\mathrm{flex}} = \sum_{j=1}^k \varphi_j \, P_j
	.
	\label{eq-reliablity-fee}
\end{equation}
Note that by construction $\sum_1^k \rho_i$ $=$ $C_f$, i.e., the sum over all
reliability payments equals the total amount of the inflexibility fees.
The quantity $P_{\mathrm{flex}}$ is the weighted sum of all available capacities,
where the weights are precisely the respective operational flexibilities.

\begin{beispiel}
	\label{bsp-toy-grid-capacity-market}%
	Assume the toy grid from Example \ref{beisp-toy-grid}.
	Then by the requirement (\ref{eq-capacity-market-flexibility-requirement})
	only three power plants can participate at the capacity market, namely
	the hydroelectric power station, the CHP plant and the gas turbine.
\begin{table*}[htp]
\centering
\begin{footnotesize}
\begin{tabular}{|l|c|r||r|r||r|r|}
	\hline
	\textbf{Power Plant} 
	 & $\mathbf{\varphi}$ 
	 & \textbf{Capacity} 
	 & \multicolumn{4}{c|}{$\mathbf{\rho_i}$ \textbf{[\euro/h]}}
	\\ \cline{4-7}
	 &
	 & \multicolumn{1}{c||}{\textbf{[MW]}}
	 & \multicolumn{2}{c||}{$\fett{C}_s$ = \textbf{205 \euro/h}}
	 & \multicolumn{2}{c|}{$\fett{C}_s$ = \textbf{790 \euro/h}}
	\\ \hline
	hydroelectric power station
		& $.980$ & $5$ \quad \ 
		&  $74$ \hspace*{.2ex} &   --- \hspace*{.5ex}
		&  $284$ \hspace*{.2ex} &   --- \hspace*{.5ex}
	\\
	gas turbine    
		& $.893$ & $5$ \quad \ 
		&  $67$ \hspace*{.2ex} &  $105$ \hspace*{.2ex}
		& $259$ \hspace*{.2ex} &  $404$ \hspace*{.2ex}
	\\
	cogeneration plant (CHP)
		& $.855$ & $5$ \quad \ 
		&  $64$ \hspace*{.2ex} &  $100$ \hspace*{.5ex}
		& $247$ \hspace*{.2ex} &  $386$ \hspace*{.2ex}
	\\
	\hline
\end{tabular}
\end{footnotesize}
\caption{\label{tab-toy-capacity-market}\footnotesize
	The three power plants participating at the capacity market of our toy grid
	in Example \ref{bsp-toy-grid-capacity-market}
	and their reliability payments $\rho$ in dependence to the total inflexibility fee
	coming from the spot market.
}
\end{table*}
	In Table \ref{tab-toy-capacity-market} they are listed with their capacities
	and the resulting reliability payments according to Equation (\ref{eq-reliablity-fee})
	and depending on the amount of total inflexibility fee coming from the
	spot market.
	For calculational details refer to the Excel file
	\url{http://math-it.org/climate/operational-flexibilities.xls}.
\end{beispiel}

\section{Discussion}
\label{sec-discussion}%
In this paper a proposal has been worked out to integrate 
operational flexibility into
the sell bids of power plants participating wholesale electricity 
spot markets.
The main idea is to calculate a fee for each power plant depending on its
operational flexibility.
For this purpose the concept of a
general measure of operational flexibility of a power plant is introduced 
here as a strictly monotonically decreasing function $\varphi$ 
of the guaranteed start-up time, normed by
condition (\ref{eq-operational-flexibility-norm-condition}).
With such a measure, the inflexibility is priced in 
by Equation (\ref{eq-price-regarding-operational-flexibility})
to the marginal price determining the sell bid
of each power plant at the spot market.
The amount depends on a market-wide reference level price $p_0$ which 
is set by the market authority or the state.
The total operational inflexibility fee $C_f$ accumulated at the 
spot markets then
is spread on the power plants participating in a given 
capacity mechanism,
depending on their operational flexibilities according to Equation
(\ref{eq-reliablity-fee}).
Here the power plants forming a capacity reserve should 
have a very high operational flexibility,
to guarantee reliability and stability of the grid.
A reasonable value is proposed by inequality 
(\ref{eq-capacity-market-flexibility-requirement}).
A simple example of a measure for operational flexibility is given 
by Equation (\ref{eq-operational-flexibility}).
Using this measure, the spot market and the corresponding 
payments to power plants participating in a capacity mechanism 
are applied to a simple but prototypical toy grid
in Examples \ref{beisp-toy-grid} and \ref{bsp-toy-grid-capacity-market}.

The most important consequence of our proposal, 
as viewed from an economic perspective,
is the internalization of the negative externality of operational 
inflexibility of power plants. 
With the inflexibility fees determined as above, the currently 
external costs would thus be paid by the spot markets 
and could be used to pay capacity reserves, 
be it on a separate capacity market or another capacity mechanism 
such as a pool of power plants forming a strategic reserve.
The inflexibility fee therefore increases welfare without necessarily
decreasing dispatch efficiency.

A critical point of our approach, however, is the determination of 
the reference level price
$p_0$. It is crucial since it can even change the 
merit order of electricity markets if it is set very high.
Although a change of the merit order in itself does not
necessarily imply severe problems,
it could nonetheless lead to the paradox that 
operationally flexible power plants participate in a short-term
spot market and therefore could not serve as a capacity reserve.
An amount $p_0$ too high would thus be adverse to the intention to 
pay a capacity mechanism and thus would even diminish welfare.
We therefore are faced with the conflicting objectives of providing 
enough means to fund the reserves of a capacity mechanism, 
and of keeping suitable power plants with high 
operational flexibility as capacity reserve.
Although this risk is calculable when choosing the amount 
for a given grid cautiously such that experiences could be 
gained over time,
a comprehensive theoretical framework to illuminate 
effects and limits of inflexibility fees 
on electricity markets should be accomplished. 
Hints to tackle this problem may be indicated 
by the optimal taxation due to Ramsey \cite{Ramsey-1927},
or by regulation theory \cite[§13]{Stroebele-et-al-2012}.
Further research in this direction appears worthwhile.


\end{document}